\documentstyle[twocolumn,aps,floats]{revtex}

\begin{document}

\draft \tolerance = 10000

\setcounter{topnumber}{1}
\renewcommand{\topfraction}{0.9}
\renewcommand{\textfraction}{0.1}
\renewcommand{\floatpagefraction}{0.9}
\newcommand{\br}{{\bf r}}

%Fixing abstract in twocolumn mode
\twocolumn[\hsize\textwidth\columnwidth\hsize\csname
@twocolumnfalse\endcsname

\title{Multifractality of time and space, covariant derivatives and gauge
invariance}
\author{L.Ya.Kobelev}
\address{Department of  Physics, Urals State University \\ Lenina
Ave., 51, Ekaterinburg 620083, Russia \\ E-mail:
leonid.kobelev@usu.ru}
 \maketitle

\begin{abstract}
A possibility to represent the standard model of fundamental
particles covariant derivatives by means of approximate
generaleralized fractional Riemann-Liouville derivatives of
multifractal time and space model is shown.
\end{abstract}

 \vspace{1cm}

%Fixing abstract in twocolumn mode
]

\section{Introduction}
1. In the Riemannian Geometry and in the standard theory of
fundamental particles (see for example, \cite{Eins},
\cite{Kane})the covariant derivatives are used, instead of usual
derivatives. In the standard theory of fundamental particles the
covariant derivatives has the form
\begin{eqnarray}\label{eq1}
&& D^\mu=\partial^\mu-ig_1 \frac{Y}{2}B^\mu-ig_2 \frac{\tau _i}{2} W_i^\mu
-ig_3 \frac{\lambda_a}{2}G_a^\mu \\ \nonumber && i=1,2,3,\quad a=1,2,..8
\end{eqnarray}
where $B^\mu$, $W_i^\mu$, $G_a^\mu$ are gauge-invariant vector
fields with symmetry groups $U(1)$,$SU(2)$,$SU(3)$,
$\tau_i$,$\lambda_a$ - are matrices of isospin and color fields,
$Y$ - hypercharge, $\mu=0,1,2,3$. The covariant derivatives with
respect to tensor $t^\mu\nu$, in space with a metric tensor
$\gamma^\mu\nu$, are defined by relations
\begin{equation}\label{eq2}
D_\alpha t^{\mu\nu}=\partial_\alpha t^{\mu\nu}+\gamma_{\alpha\beta}^\nu
t^{\mu\beta}
\end{equation}
where $\gamma_{\alpha\beta}^\nu$ are Kristoffel's coefficients
\begin{equation}\label{eq3}
\gamma_{\alpha\beta}^\nu=\frac{1}{2}\gamma^{\mu\sigma}(\partial_\alpha
\gamma_{\beta\sigma}+\partial_\beta \gamma_{\alpha\sigma}+\partial_\sigma
\gamma_{\alpha\beta})
\end{equation}
In this and other cases to $\partial^\mu$ adds vector, tensor etc.
functional terms  usual derivatives (in last cases $\partial^\mu$
is multiplied on a unit matrix of the according type).\\

2. For describing of dynamic characteristics of systems defined on
multifractal time and space sets it is necessary to define the
functionals (left-sided and right-sided) determined on the
functions, given on a multifractal sets. These functionals are the
elementary generalization of the Riemann - Liouville fractional
derivatives and integrals (about the Riemann-Liouville fractional
derivatives see \cite{Samko}) for functions defined on
multifractal sets and were introduced in \cite{Kob1}. These
functionals (generalized fractional derivatives (GFD)) read (for
differentiation with respect to time)
\begin{equation} \label{eq4}
D_{+,t}^{d}f(t)=\left( \frac{d}{dt}\right)^{n}\int_{a}^{t}
\frac{f(t^{\prime})dt^{\prime}}{\Gamma
(n-d(t^{\prime}))(t-t^{\prime})^{d(t^{\prime})-n+1}}
\end{equation}
\begin{eqnarray} \label{eq5} \nonumber
& & D_{-,t}^{d}f(t)= \\ &=& (-1)^{n}\left( \frac{d}{dt}\right)
^{n}\int_{t}^{b}\frac{f(t^{\prime})dt^{\prime}}{\Gamma
(n-d(t^{\prime}))(t^{\prime}-t)^{d(t^{\prime})-n+1}}
\end{eqnarray}
where $\Gamma$-is a gamma function, $a<b$, $a$ and $b$ stationary values
selected on an infinite axis (from $-\infty$ to $\infty$), $n-1\le d_t<n$,
$n=\{d_t\}+1$, $\{d_t\}$- integer part of $d_t \ge 0$ and $n=0$ for
$d_t<0$. Generalized fractional derivatives (GFD) (\ref{eq4})-(\ref{eq5})
coincide with fractional derivatives or fractional integrals of the
Riemann - Liouville for a case $d_t=const$. At $d_t=n+\varepsilon(t)$,
$\varepsilon(t) \to 0$ GFD are represented through usual derivatives and
integrals \cite{Kob1}. The functions in integrals in
(\ref{eq4})-(\ref{eq5}) are generalized functions, given on set of a
finitary functions \cite{Gelf}. Similar (\ref{eq4})-(\ref{eq5})
definitions GFD can be defined and with respect to space variables $\br$.
The integral functionals (\ref{eq4})-(\ref{eq5}) allow to describe dynamic
of functions defined on multifractal sets and replace for such functions
orderly or fractional (Riemann-Liouville) differentiation and integration.
These functionals partially describe the memory about past (or future, if
use right-hand GFD) time or spatial events. In \cite{Kob1} it is shown,
for $d_\alpha=1*\varepsilon_\alpha(\br(t),t)$, $\alpha=t,\br$,
$|\varepsilon| \ll 1$ where
\begin{equation}\label{eq6}
\varepsilon_\alpha=\sum\limits_i{\beta_{\alpha,i}L_{\alpha,i}(\Phi_i)}
\end{equation}
($L_{t,i}$ -are the Lagrangians densities of energy of physical fields
which presents in a point $x$, $\beta_{t,i}$ are a numerical dimension
factor, $L_{\br,i}$- are the Lagrangians densities of energy of the new
fields which appears be-cause of fractionality of space dimensions
\cite{Kob1}) that GFD can be represented in that case by usual derivatives
(scalar field):
\begin{equation}\label{eq7}
D_{+,x^\mu}^{1+\varepsilon_\alpha(x)}f(x)=\partial^\mu f(x)+a \partial^\mu
[\varepsilon_\alpha(x)f(x)]
\end{equation}
where $a$ -is a factor of the regularization of integrals
(\ref{eq4})-(\ref{eq5}) or an analogies integrals for case of
differentiation with respect to space coordinates. The relations
(\ref{eq6}), defining GFD for sets with almost integer dimension, have
more composite structure than (\ref{eq1}) and (\ref{eq2}), nevertheless,
they are very similar to definitions of the covariant derivatives. For a
gravitational field the connections GFD (for the component $\Phi_00$ of a
gravity's potentials) with covariant derivatives of effective Riemannian
space was obtained in \cite{Kob1}. The purpose of this paper is the
establishment of the connections between GFD defined by
(\ref{eq4})-(\ref{eq6}) and covariant derivatives (\ref{eq1})-(\ref{eq2}),
for taking into account only the feeble physical fields and multifractal
nature of time and using the relations for covariant derivatives
(\ref{eq4})-(\ref{eq6}) which are more complicate, than derivatives
defined by relations (\ref{eq1}) - (\ref{eq2}).

\section{Covariant derivatives and multifractal time}
1. We shall be restricted by consideration of connection GFD and covariant
derivatives defined on the  multifractal time sets (i.e. $\alpha=t$ , for
multifractal 3-dimensions coordinates space   the consideration may be
carried out by similarly methods) for a case of the small fractional
corrections to integer dimension of $t(|\varepsilon| \ll 1)$. From
(\ref{eq7}) follows
\begin{equation}\label{eq8}
D_{+,x^\mu}^{1+\varepsilon(x)}f(x)=[(1+\varepsilon)\partial^\mu-
\partial^\mu\varepsilon(x)]f(x)
\end{equation}
or, with the account of (\ref{eq6})
\begin{eqnarray}\label{eq9} \nonumber
D_{+,x^\mu}^{1+\varepsilon(x)}f(x) &=& [1-\sum\limits_i{\beta_i L_i(\Phi
_i(\br(t),t))}]\partial^\mu f(x)\\ &-& \sum\limits_i {\beta _i
\partial^\mu [L_i (\Phi_i(\br(t),t))f(x)]}
\end{eqnarray}
The relations (\ref{eq9}) defines, for $|\varepsilon| \ll 1$, connections
between GFD (in the set of multifractal time with dimension
$d=1+\varepsilon$) and orderly derivatives with respect to time and
coordinates in the time's set with dimension $d=1$. Note that from the
point of view of the multifractal theory, it is possible to treat the
time's set with dimension $d=1$, in which derivatives with respect to time
and coordinates are replaced on covariant derivatives of the (\ref{eq9}),
as the effective time's set corresponding  to multifractal time with
dimension $d=1+\varepsilon$.\\ 2. The comparison (\ref{eq9}) and
(\ref{eq1}) allows to establish difference of general structure of
derivatives (\ref{eq9}) from covariant derivatives (\ref{eq1}): except for
presence more composite, than in (\ref{eq1}), additive terms proportional
derivatives of Lagrangians density with respect to time and coordinates,
there are renormalization of the function's  factors before derivatives
$\partial^\mu$, which quantity depends both on time and from coordinates.
The presence this renormalization allows to introduce, instead of space of
time with fractional dimension for tensor fields an effective Riemannian
space with the metric defined by dependence of $\varepsilon$ from a metric
tensor (see special case in \cite{Kob1}).\\ 3. What is the physical sense
of replacement of usual derivatives on the covariant derivatives
(\ref{eq9})? The derivatives of the Lagrangians density with respect to
time and coordinates can be interpreted as a birth or disappearance of
energy (the signs of derivatives are plus or are minus, accordingly) if
the  time or the coordinates are changing. This energy is transmitted to a
field of time or is taken from it by the carrier of a measure (by the set
$R^n$). The conservation laws are fulfilled for closed system consisting
of material fields of the  time and the space and set of the carrier of a
measure $R^n$. The fields of time and space without the carrier of measure
set are open systems. The birth or annihilation of energy of the field of
time is accompanied by changes of before derivatives quantity (factor
before derivatives $\partial^\mu$) depending not only from characteristics
of function $f(t)$, with respect to which the operation of differentiation
is applied, but also from characteristics of the field of time (defined by
Lagrangians density in the given instant and given coordinates). The
comparison (\ref{eq9}) and (\ref{eq1}) allows to state if the selection of
a Lagrangians in the form of standard model of the theory of fundamental
particles (with replacement in a Lagrangians the usual derivatives by
covariant derivatives (\ref{eq1})) is made, that the covariant derivatives
(\ref{eq9}) contain the description of process of birth or disappearance
all physical fields ( for example electromagnetic fields or fields of
fundamental particles). The intensity of these processes depends on
density of energy and, thus (\ref{eq9}) are not reduced to (\ref{eq1}) at
the any selections of Lagrangians of standard model. The comparison
(\ref{eq9}) with the covariant derivatives general theory of relativity
(GTR) (\ref{eq2}) allows to state, as the components of a metric tensor
are functions of an energy-momentum tensor of gravitational field
$t_{\mu\nu}$, that covariant derivatives (\ref{eq9}) contains (\ref{eq2})
as a special case (according to presence only one gravitational field).
This case was considered  in \cite{Kob1} and it was demonstrated the
opportunity of introduction of effective Riemannian space (according  to
Riemannian space of GFR) for approximate describing of gravitational
phenomenon in multi-fractal time (if the time is multifractal in reality)
at small $\varepsilon$.

\section{Gauge invariance of the fields
$\partial_\mu\sum\limits_i{\beta_i L_i}$}

Are the fields $\partial^\mu\sum\limits_i{\beta_i L_i}$ gouge invariant?
Whether is it necessary to require the gauge invariance of GFD (for small
$\varepsilon$) from the equations of theoretical physics which are wrote
down with the help of GFD for multifractal sets, where time and space have
fractional dimensions almost undistinguished from the integer dimensions (
these equations are wrote down in \cite{Kob1})?  In the last case if to
follow fields quantum theory, the requirement of the gauge invariance
introduce the new "charges", that is the charges, defining birth or
disappearance of fields $L_i$. This charge, is defined by both factors
$\beta_i$ and factors included in Lagrangians (that define already known
charges). The requirement of the gouge invariance enters as well the new
massless physical fields: the field's of production of all known physical
fields $\varphi$:
\begin{equation}\label{eq10}
\varphi^\mu(\br(t),t),L_i)=\partial^\mu \sum{\beta _i L_i}
\end{equation}
As the relations (\ref{eq9}) for GFD are approximate and are valid only
for small $\varepsilon$, the gauge invariance (\ref{eq10}) is also
approximate and it is meaningful (if it at all is meaningful) only for
small $\varepsilon$ (though this case is most spread). The problem of
validity of introduction of a gauge invariance (invariance of derivatives
densities of Lagrangians) remains open.

\section{Is the fractal dimension of time influenced by characteristics of
multifractal sets that arise in the internal sets of "time intervals"?}
The form of the covariant derivatives in the standard theories of
fundamental particles (\ref{eq1}), allow (within the framework of
representations of multifractal time) to put the problem: is it possible
to construct dimension of time $d_{t}$ in such a manner that covariant
derivatives (\ref{eq1}) will be appears in the theory by explicit form,
instead of appears through Lagrangians? As will be shown below, it is
possible, being non-essential for $d_{t}$. It is possible presents the
covariant derivatives (\ref{eq9}) in the form more similar to covariant
derivatives (\ref{eq1}), including the relations (\ref{eq1}) as a special
case. Till now "time intervals" from which, on definition, the material
field of time is consists, were treated as "points" of time sets. It was
considered possible to neglect by interior structure of time sets
component those "time intervals". The covariant derivatives (\ref{eq1})
describe characteristics of interior symmetry of fundamental particles and
their gauge invariance, so, apparently, it is convenient in multifractal
model investigate the characteristics of inner sets that consist the time
and the space "points". So let's add to density of Lagrangians (in
$\varepsilon$ that defined the multifractal dimensions of time and space
sets)  the terms, which origin can be connected with characteristics of
sets constructing the "points" by the vicinity of points $x$. These time
and space sets ( intervals) near $x$ earlier approximately were described
as the "points" and characterized by fractional dimension $d_{t}$ or
$d_{r}$, but that part of sets contents and  carry the information about
structural characteristics of fundamental particles (as the field of time,
following \cite{Kob1}, generates all material fields and defines their
characteristics). If the sets components of  "time intervals" are
multifractal, than to each point this "interior" sets surrounding the
medial "point" with coordinate $x$ should be compared with the fractal
dimensions or with their medial integral description. Most simply in this
case for $d=1+\varepsilon(\br(t),t)$ to write down $d$ as
\begin{equation}\label{eq11}
d=1+\varepsilon=1+\sum\limits_i{\beta_i L_{0,i}(\Phi_i)}+ \sum
\limits_{i,\mu}{\int\limits_{x^\mu}^{x_0^\mu}{\tilde B_i^\mu(x)dx^\mu}}
\end{equation}
where $\tilde B^\mu$ - are vector quantifies (possessing by complicated
interior symmetry) and define characteristics of sets in a vicinity of
each point $x$ of the time (or space) intervals, in which this point is
considered. In particular, for definition of dependence $\tilde B^\mu$
from physical fields and interior symmetries of fundamental particles, for
example, relations (\ref{eq1}) can be chosen. In (\ref{eq11}) $x_0^\mu$ -
are stationary values proportional to "size" of the according  time
intervals (or space intervals), $L_{0,i}$ -are densities of Lagrangians of
free fields. As $x_0^\mu$ are very small, their contributions are
essential only for derivatives $D_{\pm,x^\mu}^{1+\varepsilon}$, which in
this case will accept the form
\begin{eqnarray}\label{eq12} \nonumber
&& D_{+,x^\mu}^{1+\varepsilon(x)}\approx[1-\sum\limits_i{\beta_i L_{0,i}
(\Phi_i(\br(t),t))]}\partial^\mu- \\ &-& \sum\limits_i {\beta _i
\partial^\mu L_{0,i}(\Phi_i(\br(t),t))}-\sum\limits_i{B_i^\mu(\br(t),t)}
\end{eqnarray}
It is possible to rewrite covariant derivatives (\ref{eq12}) as
\begin{eqnarray}\label{eq13} \nonumber
&& D_{+,x^\mu}^{1+\varepsilon(x)}\approx[1-\sum\limits_i{\beta_i L_{0,i}
(\Phi_i(\br(t),t))]}\times \\ &\times& {\partial^\mu-\frac{\sum\limits_i
{[\beta _i \partial^\mu L_{0,i}(\Phi_i(\br(t),t))-B_i^\mu(\br(t),t)]}}
{[1-\sum\limits_i {\beta_i L_{0,i}(\Phi_i(\br(t),t))}]}}
\end{eqnarray}
If   $B_i^\mu$ in (\ref{eq12})  determine by use of relations (\ref{eq1})
and if neglect by the contributions from fractal dimensions in square
brackets ($\beta \to 0$), the definition covariant derivatives
(\ref{eq12})  coincides with the covariant derivatives of standard theory
of fundamental particles (at the according selection of stationary values
and matrices). It is necessary to note, apparently, the equivalence of
mathematical expositions of dynamic characteristics of fundamental
particles in viewed model as with the help of use of densities of
Lagrangians (into which the terms describing interactions of fields and
particles are included), and exposition with the help of the introduction
of covariant derivatives. In the last case and elimination, from the
according densities of Lagrangians of terms describing interactions must
be made.

\section{Conclusions}
In the presented theory of  multifractal time and space the generalized
fractional derivatives, defined by (\ref{eq4}) - (\ref{eq5}) are used
instead of usual differentiation and integration for describing a dynamic
characteristics of any physical objects (fields, particles etc. The using
of integral functionals gives in considerable thickening of the
mathematical tool and change a great deal representations about the
physical nature of time and space. Only in the case when the fractal
corrections to FD are small, it is possible to present GFD with the help
of usual derivatives and integrals as was shown in \cite{Kob1} and used in
this paper. In this case the fractionality of dimensions of time (or,
similarly, dimensions of space, see \cite{Kob1}) can be presented by
introduction of  an "effective" time and space, in which derivatives (and
the integrals, for a case of negative fractional dimensions) are
substituted by "covariant" derivatives and integrals. The consideration of
concrete models and Lagrangians, in particular, the Lagrangians of the
standard theory of fundamental particles, has been illustrated an
opportunity of exposition of gouge invariant fields with covariant
derivatives of a new type (\ref{eq12}). This covariant derivatives are
contains, except for known terms of the gouge invariant standard theory of
fundamental particles and the terms including of effective Riemannian
space, the terms describing the continuous change of densities of energy
of physical fields (their birth or disappearance). Multifractal set of
time used in the present paper (also, as well as the mathematical tool
GFD, used in a series of other papers \cite{Kob2} - \cite{Kob4}) creates
peculiar model of the world considered as  the open system (see
\cite{Klim1},\cite{Klim2}), in which there are no invariable states. In
viewed model of multifractal time all physical fields are not stationary
(note that GFD with respect to time or coordinates of stationary values
are not equal to zero), but also their energy continuously changing thow
 potentials are invariable  ("non-potential" change of energy as the result
of an exchange of energy with the set of the carrier of a measure, because
of the presence at the covariant derivatives the terms with derivatives
from densities of Lagrangians. At least, for the case $|\varepsilon| \ll
1$, the theory of fundamental particles defined on sets of multifractal
time and space, allows to take into account by the uniform mathematical
method the influences of all known physical fields (the fundamental
particles, used by the theory, and gravitational field). It is achieved by
introduction the new paradigm \cite{Kob1}: the exposition of time and
space as multifractal sets with fractional dimensions and introduction, in
this connection, mathematical methods of generalized fractional
derivatives and integrals. For deriving the modified equations of the
standard theory of fundamental particles or various variants of the great
unification theory (or any physical theory) it is enough to describe the
fields taking into account the multifractal nature of time and space, i.e.
replace covariant derivatives such as (\ref{eq1}) (or an orderly
derivatives in a scalar theories) by generalized covariant derivatives
(GFD) (\ref{eq12}) or, depending on selection of models, by similar though
more composite covariant derivatives for fields with more complicated
mathematical nature. The problems of correspondence obtained by such
replacement models of the theories of fundamental particles to
characteristics of the real world (if use model of multifractal time and
spaces presented in \cite{Kob1}) here are not considered. All equations of
theoretical physics  wrote down with the help the GFD for the case
$\varepsilon=0$  can be considered as special cases of equations of
presented model of multifractal time and space . At last we pay attention
on the  appearance of the new characteristics in modified thus theories
(which were not explored yet) stipulated  by the additional factors, and
by the new terms in the modified covariant derivatives.

\end{document}